\documentclass[amsmat,amssymb,amsfonts,aps,prl,twocolumn]{revtex4}
\usepackage{amsmath}
\usepackage{amssymb}
\usepackage{amsthm}
\usepackage[mathscr]{euscript}
\usepackage{latexsym}
\usepackage{feynmf}
\usepackage{graphicx}
\usepackage{ulem}
\usepackage{caption}
\usepackage{algorithm}
\usepackage{algorithmic}

\newtheorem{thm}{Theorem}

\begin{document}

\title{Multiplicity of solutions to $GW$-type approximations}
\author{F. Tandetzky, J. K. Dewhurst, S. Sharma and E. K. U. Gross}
\affiliation{Max-Planck-Institut f\"{u}r Mikrostrukturphysik, Weinberg 2, D-06120 Halle, Germany}

\date{\today}


\begin{abstract}
We show that the equations underlying the $GW$ approximation have a large number of solutions. This raises the question: which is the physical solution? We provide two theorems which explain why the methods currently in use do, in fact, find the correct solution. These theorems are general enough to cover a large class of similar algorithms. An efficient algorithm for including self-consistent vertex corrections well beyond $GW$ is also described and further used in numerical validation of the two theorems. 
\end{abstract}

\maketitle
\section{Introduction}
The $GW$ approximation\cite{Aryasetiawan98} is a many-body technique used typically for the calculation of spectral density functions of solids. Its accuracy for the determination of band gaps of insulators as well as its parameter-free nature makes it a very attractive method in condensed matter physics.
The self-consistent $GW$ approximation is a fixed point method involving multidimensional objects like the Green's function and the self-energy, and it was recently demonstrated for an artificial one-point-model that this fixed point is not unique and that a different way of iterating the equations leads to a different solution\cite{Lani12}. It was further argued that including vertex corrections could exacerbate this non-uniqueness problem and lead to several solutions.
Given that the $GW$ approximation is a state of the art method for band structure calculations, such an ambiguity is a serious issue. It is also important to understand how one can go beyond $GW$ without running into unphysical solutions.

In this article we describe a new algorithm for computing the self-energy well beyond the $GW$ approximation. We investigate the nature of the additional solutions numerically and further provide general theorems, that explain why the methods currently in use to solve the $GW$ equations do indeed lead to a unique solution. Since we cover a large class of approximations these results not only validate the $GW$ calculations that are done, but they also provide conditions on approximations going beyond $GW$ for obtaining a meaningful result.

The starting point are the Hedin equations, which appear as Eqs. (A22)-(A25) in the appendix of the 1965 article of Hedin\cite{Hedin65}. We rewrite them here in modern notation:
\begin{align}
 \Gamma(1,2;3)&=\Gamma_0(1,2;3)
  +\frac{\delta\Sigma(1,2)}{\delta V(3)} \label{hed1} \\
 \Sigma(1,2)&=i\lambda\int G(1,4) W(1^+,3) \Gamma(4,2;3)d(3)d(4) \label{hed2} \\
 \Pi(1,2)&=-i\lambda\int G(2,3)G(4,2^+)\Gamma(3,4;1)d(3)d(4) \label{hed3} \\
 \frac{\delta G(1,2)}{\delta V(3)}&=\int G(1,4)G(5,2)\Gamma(4,5;3)d(4)d(5) \label{hed4} 
\end{align}
\begin{align}
 \frac{\delta W(1,2)}{\delta V(3)}&=\int W(1,4)W(5,2)\frac{\delta\Pi(4,5)}
  {\delta V(3)}d(4)d(5). \label{hed5}
\end{align}
In these equations $G$ is the Green's function, $W$ is the
renormalized Coulomb propagator\cite{note_propagator},
$\Sigma$ is the self-energy, $\Pi$ is the polarization,
$\Gamma_0(1,2;3)=\lambda\delta(1,2)\delta(1,3)$ is the bare vertex,
$\Gamma$ is the
renormalized vertex, $\lambda$ is the coupling constant,
and the potential differential $\delta V$ is the
sum of the external and Hartree contributions. The notation
$(1) \equiv x_1 \equiv ({\bf r}_1,\sigma_1,t_1)$ is used throughout\cite{note_basis}.
Note that we introduced the coupling constant $\lambda$ such that in the non-interacting case the vertex function vanishes\cite{note_basis}. Instead one could also define it in such a way that the Coulomb propagator vanishes.
Using simple transformations (introduced later in Eq. (\ref{eq_trafo}); use  $a = 1, \ b = \lambda^2 , \ c = \lambda^{-1}$) one can show that these definitions are in fact equivalent.
The physical meaning of this is, that it does not matter whether we define the non-interacting limit by switching off the interaction of the electrons with the photons or by setting the photon propagator to zero.

Hedin has shown, how one can gain an expansion of the vertex and hence of $\Sigma$ and $\Pi$ in terms of the renormalized quantities $G$ and $W$ using these equations.
That way one gets $\Sigma$ and $\Pi$ as functionals $\Sigma[G,W]$ and $\Pi[G,W]$.
In addition to the five Hedin equations are the two coupled Dyson equations:
\begin{align}\label{dyson}
\begin{split}
 G(1,2)&=G_0(1,2)+\int G_0(1,3)\Sigma(3,4)G(4,2)d(3)d(4) \\
 W(1,2)&=W_0(1,2)+\int W_0(1,3)\Pi(3,4)W(4,2)d(3)d(4),
\end{split}
\end{align}
where $G_0$ is the Green's function of the non-interacting system (which
includes the Hartree potential) and 
$W_0(1,2) = \frac{\delta(t_1-t_2)}{|\bf r_1 - \bf r_2|}$
is the bare Coulomb propagator.
Solving the Dyson equations in conjunction with the Hedin equations yields
the functionals $G[G_0,W_0]$ and $W[G_0,W_0]$.

The separation of the problem into equations for $\Sigma$ and $\Pi$ as functionals of $G$ and $W$, as well as
$G$ and $W$ as functionals of $G_0$ and $W_0$ is an important conceptual step.
In a later article by Hedin and Lundqvist\cite{Hedin69}, the equations are
combined and the functional derivative $\delta\Sigma/\delta G$ is introduced.
We would like to stress that this vertex equation together with Eqs. (\ref{hed2}), (\ref{hed3}) and the Dyson equations are not immediately useful.
One also needs the equations (\ref{hed4}) and (\ref{hed5}) in order to get an expansion of the vertex beyond $GW$.

\section{Algorithms for Hedin's equations}
Almost all practical calculations of Hedin's equations use the $GW$ approximation.
This amounts to approximating the full vertex $\Gamma$ by the bare vertex, and thus
the self-energy takes on the simple form $\Sigma(1,2)=i\lambda^2 G(1,2)W(1,2)$.
In the present work we describe a new algorithm for solving Hedin's equations which includes
corrections far beyond the $GW$ approximation. From the outset we will
insist that the computational storage requirements scale as $N^3$ and the number
of operations scale as $N^4$. These conditions are met by
Algorithm \ref{alg-hedin} (see structogram) for calculating $\Sigma[G,W]$ and $\Pi[G,W]$ using a non-trivial vertex.
Note that all the relations in the algorithm are exact
apart from the two which have the derivatives of $\Gamma$ removed -- these would
require $N^4$ storage. Solving Algorithm \ref{alg-hedin} together with the
Dyson equations we refer to as the `Starfish' algorithm. It is straight-forward to
work out which diagrams this algorithm corresponds to: finding the the
self-consistent solution to Starfish is equivalent to solving
\begin{fmffile}{fgraphs}
\begin{align}
\parbox{40\unitlength}{\centering
\begin{fmfgraph*}(30,30)
\fmfleft{l1,l2}\fmfright{r1}
\fmfrpolyn{shaded,tension=0.25,label=$\!\!\!\!\!\!\!\!\!\!\!\!\!\Gamma$}{G}{3}
\fmf{plain}{l1,G1}
\fmf{plain}{l2,G2}
\fmf{photon}{G3,r1}
\end{fmfgraph*}}
\ \ = & \qquad \quad
\parbox{40\unitlength}{\centering
\begin{fmfgraph*}(30,30)
\fmfleft{l1,l2}
\fmfright{r1}
\fmf{plain}{l1,v1}
\fmf{plain}{l2,v1}
\fmf{photon}{v1,r1}
\fmfdotn{v}{1}
\fmflabel{$\Gamma_0$}{v1}
\end{fmfgraph*}}
+
\parbox{90\unitlength}{\centering
\begin{fmfgraph*}(80,80)
\fmfleft{l1,l2}\fmfright{r1}
\fmfrpolyn{shaded,tension=0.65}{A}{3}
\fmfrpolyn{shaded,tension=0.2}{B}{3}
\fmfrpolyn{shaded,tension=0.6}{C}{3}
\fmf{plain,tension=2.5}{l2,v1}
\fmf{plain,tension=2.5}{l1,A1}
\fmf{photon,tension=2.5}{C3,r1}
\fmf{photon,tension=0.5}{A3,B1}
\fmf{photon,tension=0.55,label=$W$}{v1,v2}
\fmf{fermion,tension=0.4,label=$G$}{A2,v1}
\fmf{fermion,tension=0.0}{v2,B2}
\fmf{fermion,tension=0.6}{C2,v2}
\fmf{fermion,tension=0.6}{B3,C1}
\fmfdotn{v}{2}
\end{fmfgraph*}}  \nonumber
\\ &+
\parbox{90\unitlength}{\centering
\begin{fmfgraph}(80,80)
\fmfleft{l1,l2}\fmfright{r1}
\fmfrpolyn{shaded,tension=0.65}{A}{3}
\fmfrpolyn{shaded,tension=0.2}{B}{3}
\fmfrpolyn{shaded,tension=0.6}{C}{3}
\fmf{plain,tension=2.5}{l2,v1}
\fmf{plain,tension=2.5}{l1,A1}
\fmf{photon,tension=2.5}{C3,r1}
\fmf{photon,tension=0.5}{A3,B1}
\fmf{photon,tension=0.55}{v1,v2}
\fmf{fermion,tension=0.4}{A2,v1}
\fmf{fermion,tension=0.0}{B2,v2}
\fmf{fermion,tension=0.6}{v2,C2}
\fmf{fermion,tension=0.6}{C1,B3}
\fmfdotn{v}{2}
\end{fmfgraph}}
+
\parbox{70\unitlength}{\centering
\begin{fmfgraph}(60,60)
\fmfleft{l1,l2}\fmfright{r1}
\fmfrpolyn{shaded,tension=0.5}{A}{3}
\fmfrpolyn{shaded,tension=0.5}{B}{3}
\fmf{plain,tension=2.5}{l2,v1}
\fmf{plain,tension=2.5}{l1,A1}
\fmf{photon,tension=2.5}{B3,r1}
\fmf{photon,tension=0.5}{A2,v1}
\fmf{fermion,tension=0.5}{A3,B1}
\fmf{fermion,tension=0.5}{B2,v1}
\fmfdotn{v}{1}
\end{fmfgraph}}
\label{eq:scVertex}
\end{align}
\end{fmffile}

\noindent for the vertex, using this to find $\Sigma$ and $\Pi$ and subsequently $G$ and $W$.
The entire procedure is performed self-consistently.

\begin{algorithm}[ht]
\caption{Hedin equations solver for $\Sigma[G,W]$ and $\Pi[G,W]$.
Here the shorthand $G'(1,2;3)=\delta G(12)/\delta V(3)$, etc., is used and
the function arguments are omitted.}\label{alg-hedin}
\algsetup{indent=2em}
\begin{algorithmic}
\REQUIRE $G$ and $W$
\STATE Set $\Gamma=\Gamma_0$
\REPEAT
\STATE $G'=\int G\,G\,\Gamma$
\STATE $\Pi'=-i\lambda\int\left(G'\,G\,\Gamma+G\,G'\,\Gamma+\xout{G\,G\,\Gamma'}\right)$
\STATE $W'=\int W\,W\,\Pi'$
\STATE $\Sigma'=i\lambda\int\left(W'\,G\,\Gamma+W\,G'\,\Gamma+\xout{W\,G\,\Gamma'}\right)$
\STATE $\Gamma=\Gamma_0+\Sigma'$
\UNTIL{$\Gamma$ converged}
\STATE $\Sigma=i\lambda\int W\,G\,\Gamma$ \\
\STATE $\Pi=-i\lambda\int G\,G\,\Gamma$
\end{algorithmic}
\end{algorithm}

\section{Number and stability of solutions}
Discretization of space-time is required for the purpose of closely examining
solutions to these equations. There is some ambiguity as to how this should be done,
but at this level, we merely assume that there are $N$ space-time points in
total, and that limits in the time variable such as $G(4,2^+)$ are taken to
mean $G(4,2)$. Also the Dirac delta function in Eq. (\ref{hed1}) now becomes a
Kronenker delta function. In doing so one loses the physical meaning of these
equations, but we will assume that the true physical solution can be recovered
in the continuum limit when $N\rightarrow\infty$.

Irrespective of whether the $GW$ approximation, Starfish or some other truncation
is used, the equations to be solved form a closed system of {\it polynomial
equations} enabling us to prove general theorems about such a system.

To do so we define a concise notation. Let
\begin{align*}
 F \equiv (G,W,\Pi,\Sigma,\Gamma)\in\mathbb{C}^{n},
\end{align*}
be the vector of all dependent (or unknown) variables. Here $n=4N^2+N^3$ is the number of unknowns.
We regard the tensors $G_0$ and $W_0$ (and the bare vertex) as known and fixed.
Equations (\ref{hed2}), (\ref{hed3}), (\ref{dyson}) and a vertex equation like Eq. (\ref{eq:scVertex}) can now be written in a compact form as
$F=\mathfrak{g}(F)$ or $\mathfrak{h}(F) = 0$ with $\mathfrak{h}(F)\equiv\mathfrak{g}(F)-F$,
where $\mathfrak{g}=(\mathfrak{g}_1,\dots, \mathfrak{g}_n)$ is a set
of polynomials in several variables.

It is important to point out that most of the following considerations do not depend on the precise form of the vertex equation. For example the trivial vertex equation $\Gamma = \Gamma_0$ corresponding to the $GW$ approximation is also allowed here. In this case one can also eliminate the vertex from the equations and redefine $F, \mathfrak{g}$ and $\mathfrak{h}$ in order to include only the smaller set of quantities and equations. Similarly, if one were to fix $W$ in the Starfish algorithm then $F$ would be $(G, \Sigma, \Gamma)$ and the equations for $W$ and $\Pi$ could be eliminated.

Since we are interested in the dependence
of these equations and their solutions on the coupling strength $\lambda$, the equations to be solved are
\begin{align}\label{HeAbstract}
 F=\mathfrak{g}^{\lambda}(F) \qquad\mbox{or}\qquad \mathfrak{h}^{\lambda}(F)=0.
\end{align}

\subsection{Multiple fixed points}
\begin{figure}
\centerline{\includegraphics[width=0.9\linewidth]{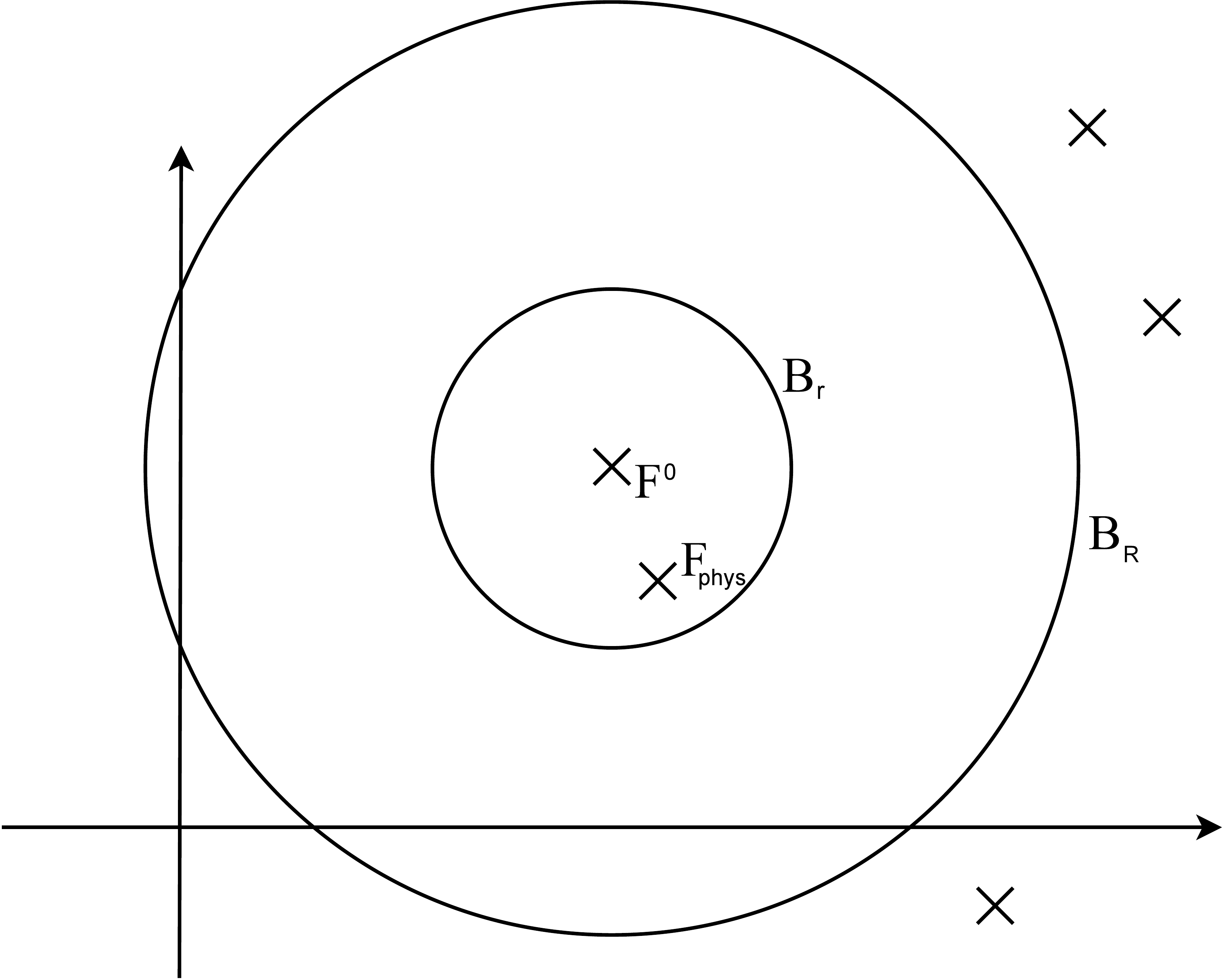}}
\caption{Sketch of the position of the solutions of the interacting system relative to the non-interacting one $F^0$. For small interaction there is one solution $F_{\rm phys}\in B_r$ of Eq. (\ref{HeAbstract}), that is close to the non-interacting one $F^0$. All others are outside $B_R$, while $r$ tends to zero and $R$ tends to infinity for the small coupling limit $\lambda\rightarrow 0$.}
\label{fixed-points}
\end{figure}

We now want to determine the number of solutions to Eq. (\ref{HeAbstract}).
An upper bound is provided by B\'{e}zout's theorem\cite{Fulton84} which states that the maximum
number of solutions to a system of polynomial equations (if finite) is equal to the product
of the total degree of each equation. Thus the $GW$ approximation with fixed
$W$ has at most $2^{N^2}$ solutions, and the Starfish algorithm, also for fixed
$W$, has at most $7^{N^3}2^{2N^2}$ solutions. The B\'{e}zout bound fails to
take into account sparsity in the polynomial equations and therefore also
degeneracy of the roots. Buchberger's algorithm is a method of systematically determining
the exact number of roots by decomposing the equations into a
Gr\"{o}bner basis\cite{Buchberger82}. This procedure is, however, computationally
very demanding and can be performed only for small (and therefore non-phyiscal)
$N$. For example, when $N=2$ the $GW$ approximation with fixed $W$ has precisely
$6$ solutions. Likewise, for $N=1$ the Starfish algorithm yields $3$ solutions.
From these considerations, it seems quite surprising that self-consistent
$GW$ works at all for realistic values of $N$. We will now provide two theorems that may explain this
apparent success.
\begin{thm}\label{thm_1}
Let $\mathfrak{h}^{\lambda}=(\mathfrak{h}_1^{\lambda},\dots,\mathfrak{h}_n^{\lambda})$,
$\lambda\in[0,1]$ be a set of polynomials $\mathfrak{h}_k^{\lambda}:\mathbb{C}^n\rightarrow\mathbb{C}$
with the following properties:
\begin{enumerate}
 \item[(i)] \label{contWrtL} They depend pointwise continuously on $\lambda$.
 \item[(ii)] \label{uniqueSol} For vanishing $\lambda$ there is exactly one solution $F^0$ to $\mathfrak{h}^0(F)=0$.
 \item[(iii)] \label{jacobianDet} The Jacobian
  $J^{\lambda}_{ij}(F)\equiv\frac{\partial\mathfrak{h}^{\lambda}_i}{\partial F_j}$
  satisfies $\det [J^0(F^0)] \ne 0$.
\end{enumerate}
Then we have 
\begin{enumerate}
 \item[(a)] For every ball $B_r$ centered at $F_0$ there is a $\lambda_0 > 0$ such that for all
  smaller $\lambda$, $\mathfrak{h}^{\lambda}$ also has a zero in $B_r$.
 \item[(b)] For every ball $B_R$ centered at $F_0$ there is a $\lambda_0 > 0$ such that for all
  smaller $\lambda$, $\mathfrak{h}^{\lambda}$ has at most one zero in $B_R$.
\end{enumerate}
\end{thm}
It can be checked easily, that these conditions are satisfied by all versions of the 
$\mathfrak{h}^{\lambda}$ we defined above.
The situation is depicted in Fig. \ref{fixed-points}. If the interaction is small,
there is only one solution $F_{\rm phys}$ that is close to the non-interacting one.
All others tend to infinity in the limit of small interaction,
i.e. for $\lambda\rightarrow 0$ one can let the radius $r$ tend to zero and $R$ tend to
infinity. This behavior suggests, that at least in some low coupling regime 
$F_{\rm phys}$ is indeed the physical solution, while all other fixed points are
far away from the right result.
\begin{proof}(a)
First we note, that owing to the continuity with respect to $\lambda$ given in
(i) and the fact that $\mathfrak{h}^{\lambda}$ is analytic then
$J^{\lambda}(F)$ also depends continuously on both $\lambda$ and $F$.
So using (iii), we can restrict $F$ and $\lambda$ to be in a region around
$F_0$ and $0$, respectively, such that $J^{\lambda}(F)$ is close enough to
$J^0(F^0)$ to have non vanishing determinant
\begin{align*}
 \det[J^{\lambda}(F)] \ne 0.
\end{align*}
We can assume, that the ball $B_r$ named in (a) was already small enough to
ensure this. Now we define
\begin{align*}
 m\equiv\min_{F\in\partial B_r}\left\{\lVert\mathfrak{h}^0(F)\rVert\right\} \ne 0,
\end{align*}
where $\partial B_r$ denotes the surface of $B_r$ and $\lVert\cdot\rVert$ denotes
the Euclidean vector norm.
Now we are interested in the minimum of $\lVert\mathfrak{h}^{\lambda}\rVert$ on
$\bar{B}_r = B_r \cup \partial B_r$. Clearly this minimum tends to zero for
$\lambda\rightarrow 0$ since $\mathfrak{h}^0(F_0)=0$. On the other hand,
the minimum of $\lVert\mathfrak{h}^{\lambda}\rVert$ with respect to $F$ on the
surface $\partial B_r$ tends to $m>0$. So we can further restrict $\lambda$
such that $\lVert\mathfrak{h}^{\lambda}\rVert$ has at least one local minimum in the interior
of $B_r$. Let's call the location of one such minimum $F_{\rm phys}$. Now that we know
that this minimum is not on the surface of $B_r$, but in the interior, we can conclude
\begin{align*}
\left. \frac{\partial\lVert\mathfrak{h}^{\lambda}\rVert^2}{\partial ({\rm Re}[F_i])} \right|_{F_{\rm phys}} = 0, \qquad
\left. \frac{\partial\lVert\mathfrak{h}^{\lambda}\rVert^2}{\partial ({\rm Im}[F_i])} \right|_{F_{\rm phys}} = 0 .
\end{align*}
This implies
\begin{align}
\left.\sum_j \frac{\partial \mathfrak{h}^{\lambda}_j}{\partial F_i} \mathfrak{h}^{\lambda*}_j\right|_{F_{\rm phys}} &= 0 \\
\sum_j  J^{\lambda}_{ji}(F_{\rm phys}) [\mathfrak{h}^{\lambda}_j(F_{\rm phys})]^* &= 0. \label{jTimesFisZero}
\end{align}
Now we have chosen $\lambda$ and the size of the ball $B_r$ such that the Jacobian
has non-zero determinant. So the only way Eq. (\ref{jTimesFisZero}) can hold is if
\begin{align*}
\mathfrak{h}^{\lambda}(F_{\rm phys}) = 0.
\end{align*}
\end{proof}
\begin{proof}(b)
We will do this proof in two steps. First we will chose an $r>0$ such that we can show
that a ball with this radius, centered at $F^0$, contains no more than one solution
(given $\lambda$ is small enough). Then we will show, that for the same ball $B_r$ and
sufficiently small $\lambda$ there is no solution in the set $B_R \setminus B_r$,
where $B_R$ is the ball named in (b).

Step 1. Define
\begin{align*}
 M^{\lambda}[q_r]\equiv \int_0^1 J^{\lambda}(q_r(s)) \, ds,
\end{align*}
with $q_r : [0,1]\rightarrow B_r $. Now $M^{\lambda}[q_r]$ can be decomposed into
$J^0(F^0)$ plus a reminder, that vanishes in the limit of $\lambda\rightarrow 0 $
and $r \rightarrow 0$. That allows us to fix $r$ and restrict $\lambda$ such that
$\det(M^{\lambda}[q_r]) \ne 0$ for any map $q_r$. Now assume we have two zeros
$F^1, F^2 \in B_r$ of $\mathfrak{h}^{\lambda}$. Then
\begin{align*}
0 &= \mathfrak{h}^{\lambda}(F^2)-\mathfrak{h}^{\lambda}(F^1) \\
  &= \int_0^1 \frac{d}{ds} \mathfrak{h}^{\lambda}[s F^2 + (1-s) F^1 ] \, ds \\
  &= \sum_j  (F^2 - F^1)_j  M^{\lambda}_{ij}[q_r],
\end{align*}
where $q_r(s)=s F^2 + (1-s) F^1$. Now we use that $\det[M]\ne 0$ and therefore
$F^2 = F^1$. So there is indeed no more than one solution in $B_r$.

Step 2. Due to the continuity with respect to $\lambda$ granted in (i) the
infimum of $\lVert\mathfrak{h}^{\lambda}\rVert$ with $F$ restricted to
$B_R \setminus B_r$ tends to the infimum of $\lVert\mathfrak{h}^0\rVert$ with the
same restriction on $F$. This is not zero, since we assumed that
$\mathfrak{h}^0(F)=0$ has only one solution $F^0$.
\end{proof}

\subsection{Convergence of iterative solutions}
In practice, one solves Eq. (\ref{HeAbstract}) iteratively. That is, one starts
with some initial guess $F_0$, inserts it into the right hand side of
$F = \mathfrak{g}(F)$ and obtains a new guess. Iterating this procedure defines a sequence
\begin{align}
 F_{i+1} = \mathfrak{g}(F_i). \label{eq:iteration}
\end{align}
A natural starting point for this is the non-interacting solution $F_0 = F^0$.
The hope is that this sequence converges to a fixed point, i.e. a solution to the equations.
{\it A priori} it is unknown if the calculation will actually converge, or if a fixed point
obtained this way actually corresponds to the {\it physical} solution. So one may wonder,
why $GW$ and similar schemes do, in fact, work in many situations.
We now show that, for weak coupling once again, a unique and convergent solution
is guaranteed.
\begin{thm}\label{thm_2}
If $\lambda$ is not too large, iterating the equations according to
(\ref{eq:iteration}), starting from the initial guess $F_0 = F^0$, converges to
the physical solution $F_{\rm phys}$.
\end{thm}
The connection to the physical solution is made through Theorem \ref{thm_1}.
\begin{proof}
For $a,b,c \in \mathbb{R}$ we define the following transformation:
\begin{align}   \label{eq_trafo}
\begin{split}
 G_0' = a G_0            \qquad W_0' = b W_0      & \qquad \Gamma_0' = c \Gamma_0   \\
 G'  = a G               \qquad W' = b W          & \qquad \Gamma' = c \Gamma \\
 \Sigma' = a^{-1}\Sigma  \qquad \Pi' = b^{-1} \Pi & \qquad \lambda' = c \lambda 
\end{split}
\end{align}
with $a^2 b c^2 = 1$.
(We will use these transformations in this section only, to avoid confusion of the
meaning of the primes with the derivatives as used earlier).
This transformation leaves Eqs. (\ref{hed2}), (\ref{hed3}), (\ref{dyson}) and e.g. (\ref{eq:scVertex}) invariant:
\begin{align*}
 F_{i+1}' = \mathfrak{g}'(F_i').
\end{align*}
We now apply the transformation with, say,
$a = \lambda^{1/4}, b= \lambda ^{1/2}, c= \lambda^{-1/2}$.
This way $\mathfrak{g}'$ depends on $\lambda$ explicitly and implicitly through
$a$, $b$ and $c$. Observe, that {\it all} coefficients appearing in $\mathfrak{g}'$
tend to zero as $\lambda\rightarrow 0$. The same is true for the transformed
starting values $F_0' = {F^0}'$. In this situation Banach's fixed point theorem
can be applied for the map $\mathfrak{g}'$ defined in a vicinity of the non-interacting
solution ${F^0}'$. We conclude, that for small $\lambda$ the transformed quantities
tend to a fixed point. Since the transformation can be inverted, this remains true
for the original quantities. By Theorem \ref{thm_1} for small $\lambda$ the
solution $F_{\rm phys}$ is the solution nearest to the non-interacting one.
Hence the fixed point obtained is indeed $F_{\rm phys}$.
\end{proof}

\subsection{Generalizations of the Theorems}

For the sake of clarity, we presented the Theorems \ref{thm_1} and \ref{thm_2} slightly less general than possible. However, these Theorems can be easily generalized.
For example in the proof of Theorem \ref{thm_1} there was no reference to the fact that we are dealing with polynomials, rather only to the preposition that the functions are analytic in the region of interest (i.e. $\bar B_r$ and $\bar B_R$). A consequence of this is that Theorem \ref{thm_1} also applies to vertex equations that are more sophisticated than just polynomials.
It also allows us to use Dyson's equations in the `solved' form, e.g. $G=(1-G_0 \Sigma)^{-1} G_0$, if we restrict $\Sigma$ and $\Pi$ (and hence $B_R$ and $B_r$) such that the introduced poles are outside $\bar B_R$ and $\bar B_r$. This restriction is natural in the non-interacting limit since then the self-energy and polarization should tend to zero.
Similarly, one can extend the proof of Theorem \ref{thm_2} to these solved Dyson equations (note that they still obey the symmetry used in the proof). Also the restriction of choosing the non-interacting solution as starting point is actually not necessary, though it renders the proof much simpler; it can be shown that the fixed point becomes attractive in the small coupling limit, even without the restriction of starting from zero self-energy and polarization.

\section{Numerical investigation}

\begin{figure}[ht]
\centerline{\includegraphics[width=0.9\linewidth]{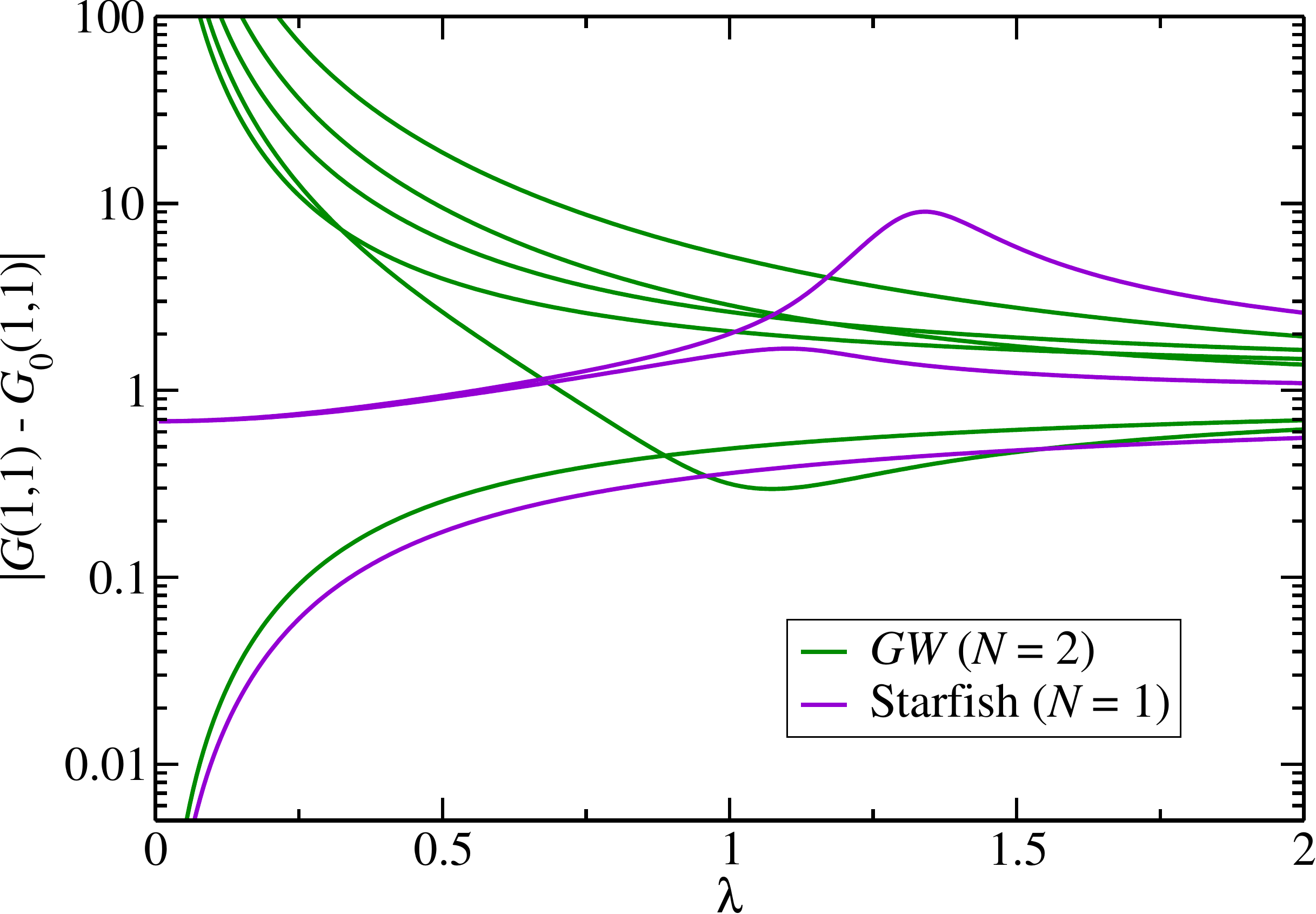}}
\caption{Plot of the distance of a matrix element of the Green's function to the non-interacting one versus the coupling strength $\lambda$ for all possible solutions of $GW$ with $N=2$ and Starfish with $N=1$, for random $G_0$ and $W$. $W$ is kept fixed in both cases. Always only one solution tends to the non-interacting one for the weak coupling limit.}
\label{GW+Starfish}
\end{figure}

Numerical checks of the above theorems were performed for both self-consistent
$GW$ and the Starfish algorithm. As already mentioned, obtaining all solutions is in practice only possible for very small $N$. Hence for $GW$ we set $N=2$ and for Starfish we use $N=1$.
Plotted in Fig. \ref{GW+Starfish} are
{\it all} the solutions for these algorithms, as a function of
$\lambda$. The numerical input,
in this case $G_0$ and $W$, were chosen to be random complex numbers, and
$W$ was kept fixed (which is common practice for real $GW$ calculations). As mentioned
earlier, there are 6 solutions in the $GW$ case. Of these, 5 tend to infinity
and the remaining solution tends to $G_0$ as $\lambda\rightarrow 0$. This is
a visualization of Theorem \ref{thm_1}. For Starfish, 2 of the 3 solutions
tend to a constant. This may seem to be in violation of the theorem, but
in this case the vertex $\Gamma$ (and therefore $F$) diverges.

\begin{figure}[ht]
\centerline{\includegraphics[width=0.9\linewidth]{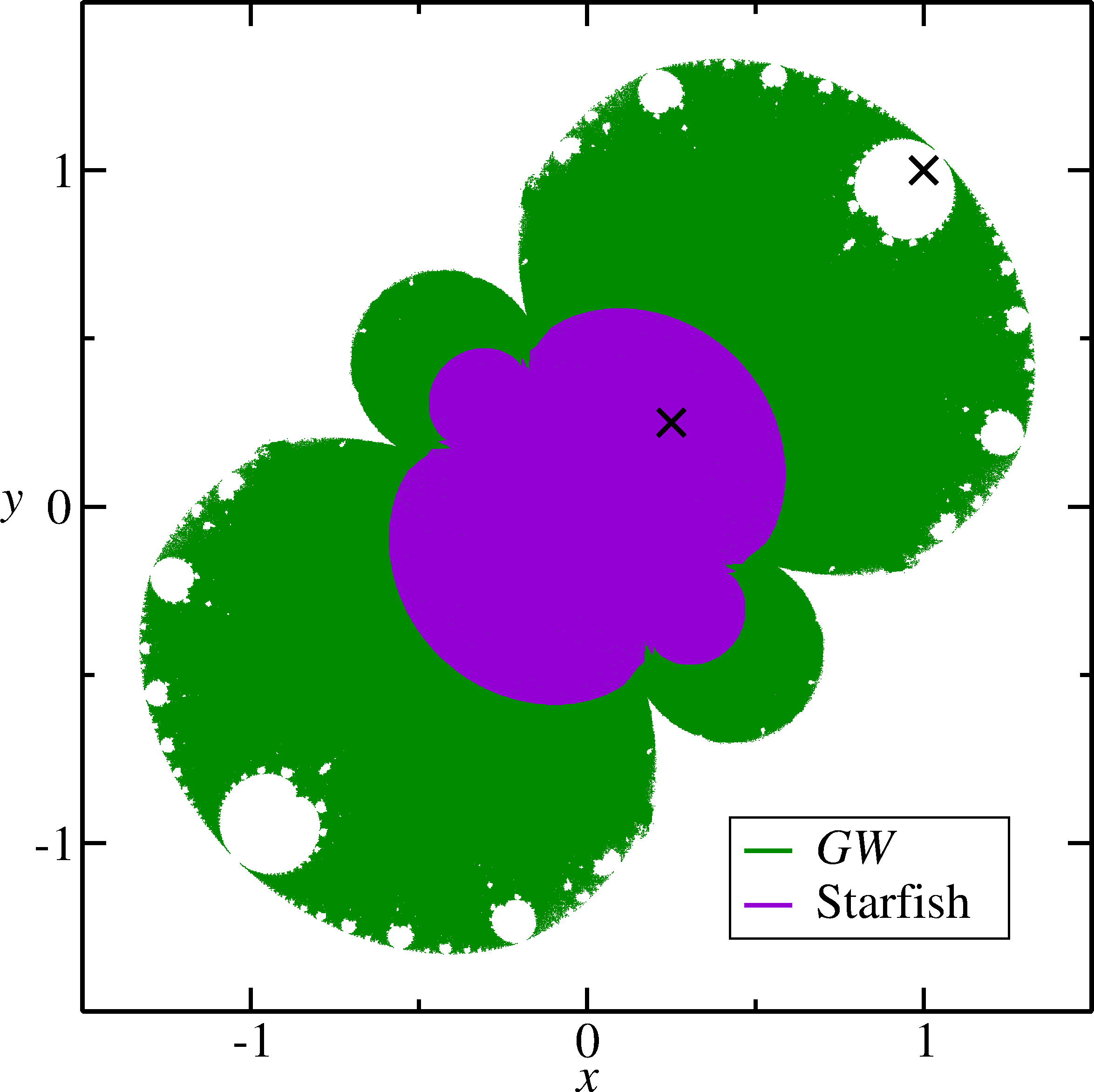}}
\caption{Domain of convergence of $GW$ and Starfish with $N=1$
for input values of
$G_0$ in the complex plane, when using the non-interacting solution as starting point. Here $W_0=1$ and $\lambda=1$. The crosses mark the chosen $G_0$ for investigating the starting point dependence while fixing $G_0$, see Fig. \ref{julia}.}
\label{mandelbrot}
\end{figure}

\begin{figure}[ht]
\centerline{\includegraphics[width=0.9\linewidth]{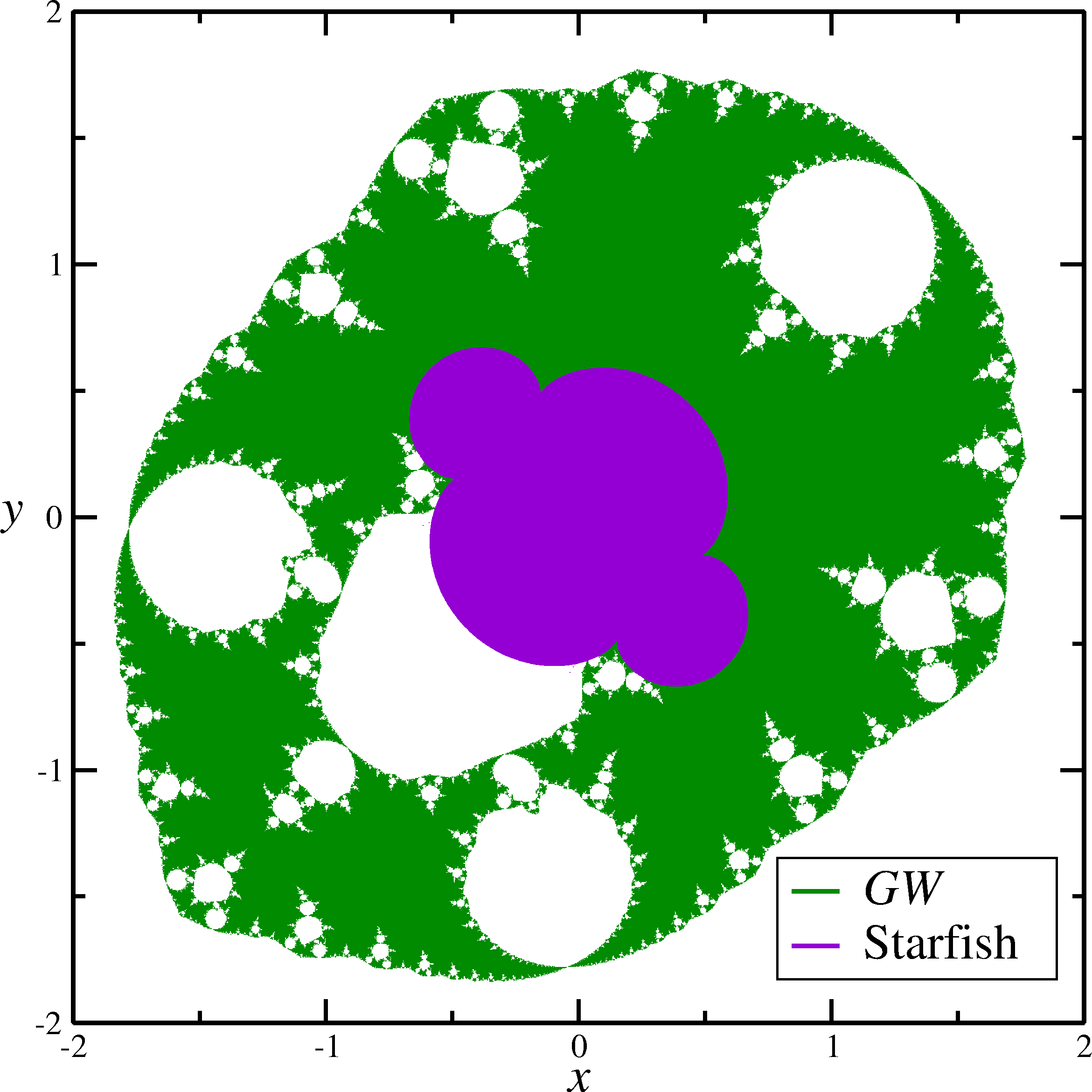}}
\caption{Domain of convergence of $GW$ and Starfish with $N=1$ for different starting points
of the fixed point cycle. The values of $G_0$ are fixed to $1+i$ for $GW$ and
$1/4+i/4$ for Starfish, as indicated by the
crosses in Fig. \ref{mandelbrot}.}
\label{julia}
\end{figure}

We can also examine the domains of convergence for both of these algorithms.
For $N=1$ we fix $W_0=1$ and $\lambda=1$ and plot the region of convergence
of $G_0$ for the fully self-consistent $GW$ and Starfish algorithm (for which
$W$ is also computed self-consistently) in Fig. \ref{mandelbrot}.
It can be observed that the region of
stability shrinks for the higher-order method. Also noteworthy is that the
region has a fractal boundary (this may be unsurprising since for case
$G_0W=1$ the domain is the Mandelbrot set).
Perhaps more interesting is the region of {\it starting points} for which
the algorithms converge. These are plotted in Fig. \ref{julia} for
the same $W_0$ and $\lambda$ but this time with $G_0=1+i$ for $GW$ and
$G_0=1/4+i/4$ for Starfish, and with a variable starting point for $G$.
Once again the region of convergence is smaller for Starfish, but in both cases
{\it only one} solution is found, irrespective of the starting point.
This is a numerical confirmation of Theorem \ref{thm_2}. Note that for $GW$ a situation was picked, where the non-interacting starting point does {\it not} lead to convergence. Hence this can be considered a large coupling situation. But still there seems to be only one stable fixed point.
The boundary of the region is also fractal (this corresponds to the Julia set).

\section{Conclusions}
We have argued that truncating Hedin's equations to some order yields
systems of polynomial equations which have a very large number of solutions.
As an example of this, the Starfish algorithm was introduced which includes
vertex corrections beyond $GW$ and consequently has even more fixed point solutions.
The number of solutions tends to infinity as either the order of truncation or $N$ tends to infinity,
reflecting the inherent problem of solving Hedin's equations as a functional
differential equation.
Two theorems were presented that shed some light on the general behavior of these fixed points.
In particular we have shown, that there is exactly one solution that tends to the
non-interacting case for small coupling, while all others are divergent in this limit.
Numerical tests of self-consistent $GW$ and the Starfish algorithm for small $N$
demonstrated that the system also converges uniquely to one fixed point
even for fairly large coupling. Furthermore, the region of stability may be fractal in
nature, indicating that finding simple necessary and sufficient conditions for
ensuring convergence of $GW$ calculations {\it a priori}, may be impossible.

\acknowledgments{We thank Lucia Reining, Martin Stankovski and Ralph Tandetzky for useful
discussions.}

\bibliographystyle{apsrmp}
\bibliographystyle{unsrt}
\bibliography{article}

\begin{thebibliography}{1}
\bibitem{note_propagator}
We stress that $W$ and $W_0$ are two-point objects contrary to the interactions
  being four point objects.
\bibitem{note_basis}
Introducing two more integrations and using the bare vertex one can rewrite
  Eqs. (\ref{hed2}), (\ref{hed3}) such that they involve tensor contractions
  only. That way, all our equations of interest Eqs.
  (\ref{hed1})-(\ref{eq:scVertex}) are basis-independent, though the bare
  vertex becomes more complicated in a different basis. Also the coupling
  constant $\lambda$ no longer appears anywhere else than in the definition of
  the bare vertex $\Gamma_0$.
\end{thebibliography}


\end{document}